\documentclass[twocolumn,showpacs,preprintnumbers,amsmath,amssymb]{revtex4}
\usepackage{graphicx}
\usepackage{dcolumn}
\usepackage{bm}
\begin{document}

\title{Multipartite entanglement for entanglement teleportation}

\author{Jinhyoung Lee}

\email{jlee@am.qub.ac.uk}

\affiliation{School of Mathematics and Physics, The Queen's University,
  Belfast, BT7 1NN, United Kingdom}

\affiliation{Institute of Quantum Information Processing and Systems,
  University of Seoul, Seoul, Korea}

\author{Hyegeun Min}

\affiliation{Department of Physics, Sookmyung Women's University, Seoul,
  Korea}

\author{Sung Dahm Oh}

\email{sdoh@sookmyung.ac.kr}

\affiliation{Department of Physics, Sookmyung Women's University, Seoul,
  Korea}

\date{\today}

\begin{abstract}
  The scheme for entanglement teleportation is proposed to incorporate
  multipartite entanglement of four qubits as a quantum channel. Based
  on the invariance of entanglement teleportation under arbitrary
  two-qubit unitary transformation, we derive relations of
  separabilities for joint measurements at a sending station and for
  unitary operations at a receiving station. From the relations of
  separabilities it is found that an inseparable quantum channel always
  leads to a total teleportation of entanglement with an inseparable
  joint measurement and/or a nonlocal unitary operation.
\end{abstract}

\pacs{03.67.-a, 03.67.Hk}

\maketitle

\section{Introduction}

Quantum teleportation is one of the most striking features emerging from
quantum entanglement which is inherent in quantum mechanics
\cite{Bennett93}.  Entangled systems divided into two parts enable to
transfer the quantum information of an unknown quantum state to a remote
place while the original state is destroyed.  No information of the
unknown state is ever revealed during teleportation process.  Quantum
teleportation has been especially interested in single body systems of
two-level, $N$-dimensional, and continuous variable states
\cite{Bennett93,Stenholm98,Vaidman94}.

Entanglement teleportation is to transfer the entanglement initially
imposed on an unknown multipartite state to a multipartite state at a
remote place \cite{Lee00}. The entanglement is transferred onto the
composite system of subsystems which have never directly interacted. In
this sense, entanglement teleportation is similar to entanglement
swapping \cite{Zukowski93}.  However, entanglement teleportation
transfers not only the amount of entanglement but also the entanglement
structure (entangled state itself). Entanglement teleportation of two
qubits has recently been studied for pure and noisy quantum channels
\cite{Lee00,Ikram00}. It is closely related to quantum computation as
the two-qubit teleportation together with one-qubit unitary operations
is sufficient to implement universal gates required for quantum
computation \cite{Gottesman99}.

In the earlier protocols for two-qubit teleportation, separate
Einstein-Podolsky-Rosen (EPR) pairs are utilized for the quantum channel
so that the joint measurement is decomposable into two independent
Bell-state measurements and the unitary operation into two local
one-qubit operations. This implies that entanglement teleportation can
be implemented by a series of single-qubit teleportations
\cite{Bennett93,Lee00} which we call ``a series teleportation of
entanglement''. It is desirable to ask the following questions: whether
a quantum channel is restricted only to EPR entanglement, if not, what
other types of entanglement are possible, and how they play a role in
entanglement teleportation. These questions have been addressed in part
by employing Greenberger-Horne-Zeilinger (GHZ) entanglement
\cite{Greenberger89} of three and four qubits as a quantum channel
\cite{Gorbachev00, Lee01}. However, the investigations have been
restricted thus far to partially unknown entangled states such as $a
|01\rangle + b |10\rangle$ and do not cover all possible states of a
two-qubit system.

In this paper we consider entanglement teleportation of completely
unknown entangled states such as
\begin{equation}
  \label{eq:pus}
  a |00\rangle + b |01\rangle + c |10\rangle + d |11\rangle
\end{equation}
where $a$, $b$, $c$, and $d$ are complex numbers and $\{|ij\rangle\}$ is
an orthonormal basis set. The present scheme is formulated so as to
employ multipartite entanglement of four qubits as a quantum channel;
the composite system of four qubits may have various types of
entanglement, for example, two EPR pairs, four GHZ triads, etc. We show
that entanglement teleportation has an invariance under arbitrary
two-qubit unitary transformation and variant protocols are available. By
the invariance of entanglement teleportation, we derive relations of
separabilities for joint measurements at a sending stations and for
unitary operations at a receiving station. Due to the relations of
separabilities, we show that an an inseparable quantum channel always
leads to a ``total teleportation of entanglement'', which employs an
inseparable joint measurement and/or a nonlocal unitary operation, as
opposed to a series teleportation of entanglement.

\begin{figure}[htbp]
  \centering

  \caption{A schematic drawing of a total teleportation of two-qubit
    unknown state (\ref{eq:us}) via an inseparable quantum channel of
    four qubits $A_1$, $A_2$, $B_1$, and $B_2$. The inseparable quantum
    channel is in the state given by Eq.~(\ref{eq:meqc}) as the two EPR
    pairs generated in the entangler are transformed by the nonlocal
    unitary operator $\hat{V}\hat{U}^T$ in the two-qubit gate. Suppose
    Alice obtains an outcome $(\alpha,\beta)$ in her joint measurement
    represented by the basis set $\{|\mu_{\alpha\beta}\rangle\}$.  After
    receiving the four-bit classical message $(\alpha,\beta)$, Bob
    applies the corresponding unitary operator $\hat{U}_{\alpha\beta}$.}
  \label{fig:1}
\end{figure} 

\section{\label{sec:mqt} two-qubit teleportation}

In the original proposal \cite{Bennett93}, quantum teleportation
utilizes an EPR pair as a quantum channel which is shared by a sender,
Alice, and a receiver, Bob.  After she receives a particle in an unknown
state and one of the entangled pair, Alice performs a joint measurement
on their composite state. She transmits the outcome to Bob through a
classical channel. Bob applies a suitable unitary operation on his
particle of the entangled pair, which is chosen in accordance with the
outcome of the joint measurement. The final state of Bob's particle is
completely equivalent to the original unknown state if the quantum
channel is maximally entangled.

A completely unknown state of two qubits $U_1$ and $U_2$ to teleport can
be represented by
\begin{eqnarray}
    \label{eq:us}
    |\phi_{u}\rangle_{U} &=&
    \sum_{i,j=0}^{1}c_{ij}|i,j\rangle_U 
\end{eqnarray}
where the subscript $U$ denotes the composite system of two qubits $U_1$
and $U_2$, $c_{ij}$ is a complex number and $\{|i,j\rangle_{U}\}$ is an
orthonormal basis set; $|i,j\rangle_{U} = |i\rangle_{U_1} \otimes
|j\rangle_{U_2}$ with the basis set $\{|0\rangle_q, |1\rangle_q\}$ of
qubit $q$. Note that the unknown state in Eq.~(\ref{eq:us}) is entangled
unless the coefficient matrix $c_{ij}$ is decomposable such that $c_{ij}
= d_{i}e_{j}$ for some complex vectors $d_{i}$ and $e_{j}$.

We consider a quantum channel of four qubits which are divided into two
parts, {\em i.e.}, two qubits are sent to Alice and the others to Bob as
shown in Fig. \ref{fig:1}. Alice's two qubits $A_1$ and $A_2$ are
denoted by $A$ and Bob's two qubits $B_1$ and $B_2$ by $B$. A perfect
teleportation requires that two parts of $A$ and $B$ be in a maximally
entangled pure state, that is, the state $|\phi_c\rangle_{AB}$ of the
quantum channel satisfies the relation,
\begin{eqnarray}
\label{eq:ido}
\mbox{Tr}_{B(A)}\left(|\phi_c\rangle_{AB}\langle\phi_c|\right) =
\frac{1}{4} \openone_{A(B)}, 
\end{eqnarray}
where $\mbox{Tr}_{i}$ is a partial trace over subsystem $i$ and
$\openone_i$ an identity operator of part $i$. The channel state
$|\phi_c\rangle$ can be written by Schmidt decomposition as
\begin{eqnarray}
\label{eq:meqc}
 |\phi_c\rangle_{AB} &=& \frac{1}{2}
 \sum_{i,j=0}^{1}
 |\psi_{ij}\rangle_A \otimes |\varphi_{ij}\rangle_B, 
\end{eqnarray}
where $|\psi_{ij} \rangle_A = \hat{U} |i,j\rangle_A$, $|\varphi_{ij}
\rangle_B = \hat{V} |i,j\rangle_B$, and $\hat{U}$ and $\hat{V}$ are
two-qubit unitary operators. Note that $|\psi_{ij} \rangle$ is an
entangled state of Alice's two qubits if the unitary operator $\hat{U}$
is nonlocal. Similarly, $|\varphi_{ij} \rangle$ is an entangled state of
Bob's two qubits if $\hat{V}$ is nonlocal.

The channel state $|\phi_c\rangle_{AB}$ in Eq.~(\ref{eq:meqc}) can be
represented in the more convenient form of
\begin{eqnarray}
  \label{eq:qcsf}
 |\phi_c\rangle_{AB}=(\openone_A \otimes
\hat{V}\hat{U}^T) |\bar{\phi}_c\rangle_{AB}
\end{eqnarray}
where
\begin{eqnarray}
   \label{eq:mqc2}
   |\bar{\phi}_c\rangle_{AB} &=& \frac{1}{2}\sum_{i,j=0}^{1}
   |i,j\rangle_{A} \otimes |i,j\rangle_{B}.
\end{eqnarray}
The state $|\bar{\phi}_c\rangle$ is also a maximally entangled state of
the two parts $A$ and $B$.  On the other hand, it is separable in
$(A_1,B_1)$ and $(A_2,B_2)$ such that $|\bar{\phi}_c\rangle_{AB} =
|\mbox{EPR}\rangle_{A_1B_1} \otimes |\mbox{EPR}\rangle_{A_2B_2}$ where
$|\mbox{EPR}\rangle=\sum_i |i,i\rangle/\sqrt{2}$. The state
$|\bar{\phi}_c\rangle$ has been used as a quantum channel for
entanglement teleportation \cite{Lee00,Ikram00}.

A pure generalized GHZ state of four qubits is defined by using
generalized Schmidt decomposition as \cite{Thapliyal99}
\begin{eqnarray}
  \label{eq:nsps}
  |\phi_4\rangle_{AB} = \sum_{i=0}^{1} \lambda_i |\alpha_i\rangle_{A_1}
   \otimes |\beta_i\rangle_{A_2} \otimes |\gamma_i\rangle_{B_1}
   \otimes |\delta_i\rangle_{B_2}
\end{eqnarray}
where $\{|\alpha_i\rangle\}$, $\{|\beta_i\rangle\}$,
$\{|\gamma_i\rangle\}$, and $\{|\delta_i\rangle\}$ are orthonormal
vector sets and $\lambda_i$'s are positive. A pure generalized GHZ state
is not a good candidate for a quantum channel because it does not
fulfill the requirement~(\ref{eq:ido}) of maximal entanglement of two
parts $A$ and $B$. More explicitly,
\begin{eqnarray}
  \label{eq:qsps}
  \mbox{Tr}_{B}\left(|\phi_4\rangle_{AB}\langle\phi_4|\right) =
  \sum_{i=0}^1 |\lambda_i|^2
  |\alpha_i,\beta_i\rangle_{A}\langle\alpha_i,\beta_i|, 
\end{eqnarray}
which is not proportional to $\openone_A$. In fact it is proportional to
a projector which projects a state into a subspace spanned by
$\{|\alpha_i,\beta_i\rangle_{A}\}$. We note however that a single-qubit
teleportation can be performed via a quantum channel of three qubits
which is in a maximal GHZ state \cite{Karlsson98} and thus there could
be a possibility that a GHZ state of more than two times of the
teleporting qubits may lead to a perfect teleportation.

Alice performs a joint measurement on the four qubits, $A_1$, $A_2$,
$U_1$, and $U_2$.  The joint measurement is constructed using a set of
sixteen projectors
$\{\hat{M}_{\alpha\beta}=|\mu_{\alpha\beta}\rangle_{AU}
\langle\mu_{\alpha\beta}|\}$ where
\begin{eqnarray}
  \label{eq:jmp}
  |\mu_{\alpha\beta}\rangle_{AU} =
   \left(\openone_A\otimes\hat{U}_{\alpha\beta}\right)
   |\phi_c\rangle_{AU}.
\end{eqnarray}
Here $|\phi_c\rangle_{AU}$ is the same as the state given in
Eq.~(\ref{eq:qcsf}) and
$\hat{U}_{\alpha\beta}=\hat{\sigma}_\alpha\otimes \hat{\sigma}_\beta$ is
a local unitary operator with Pauli spin operators
$\hat{\sigma}_\alpha=\openone$, $\hat{\sigma}_x$, $\hat{\sigma}_y$, and
$\hat{\sigma}_z$. The set $\{\hat{M}_{\alpha\beta}\}$ satisfies a
completeness relation as
\begin{eqnarray}
  \label{eq:jmpcr}
  \sum_{\alpha,\beta=1}^4 \hat{M}_{\alpha\beta} = \openone_{A} \otimes \openone_{U}.
%|\mu_{\alpha\beta}\rangle_{AU}
%\langle\mu_{\alpha\beta}| &=& \openone_{A}
%  \otimes \openone_{U}.
\end{eqnarray}
Further the sixteen projectors are orthogonal such that
\begin{eqnarray}
  \label{eq:orth}
  \hat{M}_{\alpha\beta} \hat{M}_{\gamma\delta} =
  \mbox{Tr}\left(
  \hat{U}^\dagger_{\alpha\beta}\hat{U}_{\gamma\delta} \right)
  |\mu_{\gamma\delta}\rangle_{AU}\langle\mu_{\alpha\beta}| = 
\delta_{\alpha\gamma} \delta_{\beta\delta} \hat{M}_{\alpha\beta}.
\end{eqnarray}
This implies that the joint measurement represented by
$\{\hat{M}_{\alpha\beta}\}$ is an orthogonal measurement on the
composite system of $A$ and $U$.

A key step is to evaluate a partial inner product ${}_{AU}\langle
\mu_{\alpha\beta}|\phi_c\rangle_{AB}$ by applying an identity operator
$\openone_U = \sum_{i,j} |i,j\rangle_U \langle i,j|$ on the right side:
\begin{eqnarray}
  \label{eq:twist}
  {}_{AU}\langle \mu_{\alpha\beta}|\phi_c\rangle_{AB} &=& \frac{1}{4}
  \hat{U}_{\alpha\beta}^\dagger \hat{{\cal T}}_{BU},
\end{eqnarray}
where $\hat{{\cal T}}_{BU}=\sum_{i,j} |i,j\rangle_B {}_U\langle i,j|$ is
a transfer operator from a state of $U$ to that of $B$ such that
$\hat{{\cal T}}_{BU} |\phi\rangle_U = |\phi\rangle_B$. The form of
$\hat{U}^\dagger \hat{{\cal T}}$ plays a crucial role in revealing an
invariance of entanglement teleportation which will be discussed in the
next section.

The state $|\Psi\rangle_{UAB}$ of the whole composite system of $U$,
$A$, and $B$ can be represented with respect to the basis set
$\{|\mu_{\alpha\beta}\rangle_{AU}\}$ of the joint measurement as follows
\begin{eqnarray}
|\Psi\rangle_{UAB} &=& |\phi_u\rangle_U\otimes|\phi_c\rangle_{AB}
\nonumber \\
&=& \left(\sum_{\alpha,\beta=1}^4 \hat{M}_{\alpha\beta} \right)
|\phi_c\rangle_{AB}\otimes|\phi_u\rangle_U \nonumber \\ 
&=& \frac{1}{4}
\sum_{\alpha,\beta=1}^{4}|\mu_{\alpha\beta}\rangle_{AU}\otimes
\hat{U}_{\alpha\beta}^\dagger \hat{{\cal T}}_{BU} |\phi_u\rangle_U.
\end{eqnarray}
Suppose Alice obtains an outcome $(\alpha,\beta)$ when she performs the
joint measurement on the composite system of $A$ and $U$. Bob's two
qubits come to be in the state of
$\hat{U}_{\alpha\beta}^\dagger|\phi_u\rangle_B$.  When he receives
through a classical communication the four-bit message concerning the
outcome $(\alpha,\beta)$, Bob applies the corresponding unitary
operation $\hat{U}_{\alpha\beta}$ on his qubits, which completes the
two-qubit teleportation process.

\section{relations of separabilities for joint measurements and for unitary
  operations}
\label{sec:dju} 

In the proposed protocol of two-qubit teleportation, we employed an
orthogonal measurement for the joint measurement. We may consider a
positive operator valued measurement for a joint measurement,
such that for a set of unitary operators $\{\hat{U}_g\}$ with the order
$G$
\begin{eqnarray}
  \label{eq:uocr}
  \frac{1}{G}\sum_{g} \openone \otimes \hat{U}_g
|\phi\rangle_{AU} \langle\phi| \openone \otimes
\hat{U}_g^\dagger = \frac{1}{4^2} \openone_A \otimes \openone_U
\end{eqnarray}
where $|\phi\rangle_{AU}$ is a maximally entangled state of $A$ and $U$.
This type of a positive operator valued measurement was studied for
universal teleportation \cite{Braunstein00}. If $\hat{U}_g =
\hat{\sigma}_{\alpha}\otimes\hat{\sigma}_\beta$, this measurement is
simply equal to the orthogonal joint measurement represented by the
bases in Eq.~(\ref{eq:jmp}).

We shall show an invariance of entanglement teleportation under
arbitrary two-qubit unitary transformation. For a maximally entangled
state $|\phi\rangle$ of two parts, let $\{|\mu_g\rangle_{AU} =
\openone_A \otimes \hat{U}_g |\phi\rangle_{AU}\}$ be a set of joint
measurement bases and $\{|\phi_{g'}\rangle_{AB} = \openone_A \otimes
\hat{U}_{g'} |\phi\rangle_{AB}\}$ be a set of unitarily transformed
channel states.  The partial inner product of $|\mu_g\rangle_{AU}$ and
$|\phi_{g'}\rangle_{AB}$ is obtained as
\begin{eqnarray}
  \label{eq:jud}
  {}_{AU}\langle \mu_g|\phi_{g'}\rangle_{AB} &=& \frac{1}{4}
  \hat{U}_{g'}\hat{U}_g^\dagger {\cal T}_{BU}.
\end{eqnarray}
When $g=g'$, this is just a transfer operator. The teleportation is
completely specified by $G$ pairs of joint measurement bases and their
corresponding channel states, $\{|\mu_g\rangle, |\phi_g\rangle\}$.
The partial inner product in Eq.~(\ref{eq:jud}) is invariant under the
transformation of
\begin{eqnarray}
  \label{eq:dtju1}
|\mu_g\rangle_{AU} &\rightarrow& \hat{W}^T_r \otimes \hat{W}_l
|\mu_g\rangle_{AU}
\end{eqnarray}
and
\begin{eqnarray}
  \label{eq:dtju2}
|\phi_g\rangle_{AB} &\rightarrow& \hat{W}^T_r \otimes \hat{W}_l
|\phi_g\rangle_{AB},
\end{eqnarray}
for each $g$ with some two-qubit unitary operators $\hat{W}_l$ and
$\hat{W}_r$. Thus one may have variant protocols of entanglement
teleportation under the transformation in Eqs.~(\ref{eq:dtju1}) and
(\ref{eq:dtju2}), due to the arbitrariness of $\hat{W}_l$ and
$\hat{W}_r$. We note here that the invariance of entanglement
teleportation may be extended further with respect to a rather general
completely positive operation \cite{Jamiolkowski72}.

The invariance of entanglement teleportation raises relations of
separabilities for joint measurements and for unitary operations. In
particular, an inseparable joint measurement may be transformed into two
independent Bell-state measurements and/or a nonlocal unitary operation
into a local operation.  A joint measurement is said to be separable
when each measurement basis can be decomposed into a product state of
either $(A_1,U_1)$ and $(A_2,U_2)$ or $(A_1,U_2)$ and $(A_2,U_1)$.
Further a protocol of entanglement teleportation is called a series
teleportation of entanglement when its joint measurement is separable
{\em and} the corresponding unitary operation is local. The series
teleportation of entanglement consists of independent Bell-state
measurements and local unitary operations \cite{Lee00,Ikram00}.
Otherwise, it is called a total teleportation of entanglement in the
sense that it is not decomposable into a series of single-qubit
teleportation \cite{Lee00}.

In Sec.~\ref{sec:mqt} we presented a protocol of total teleportation of
entanglement with an inseparable joint measurement and a local unitary
operation when the quantum channel state in Eq.~(\ref{eq:qcsf}) is
inseparable.  One may try to construct a series teleportation of
entanglement by using the invariance of entanglement teleportation under
the transformation of Eqs.~(\ref{eq:dtju1}) and (\ref{eq:dtju2}).
Suppose that a joint measurement becomes separable in $(A_1,U_1)$ and
$(A_2,U_2)$ for some $\hat{W}_l$ and $\hat{W}_r$ such that
\begin{eqnarray}
  \label{eq:tjmo}
  |\mu_{\alpha\beta}\rangle_{AU} \rightarrow
   |\tilde{\mu}_{\alpha\beta}\rangle_{AU} = \openone_A \otimes
   \bar{U}_{\alpha\beta}|\bar{\phi}_c\rangle_{AU} 
\end{eqnarray}
where $\bar{U}_{\alpha\beta}=\hat{W}_l \hat{U}_{\alpha\beta} \hat{V}
\hat{U}^T \hat{W}_r = \sigma_\alpha \otimes \sigma_\beta$. Then, the
corresponding unitary operators are transformed as
\begin{eqnarray}
  \label{eq:tuo}
  \hat{U}_{\alpha\beta} \rightarrow \bar{U}_{\alpha\beta} (\hat{V}\hat{U}^T)^\dagger.
\end{eqnarray}
The transformed unitary operators are clearly nonlocal since $\hat{V}
\hat{U}^T$ is nonlocal due to the inseparability of $|\phi_c\rangle$.
The new protocol consists of the separable joint measurement and the
nonlocal unitary operation, which is the opposite case to the
untransformed protocol of the inseparable joint measurement and local
unitary operation. However, the altered protocol is a total
teleportation as well. An inseparable quantum channel always leads to a
total teleportation of entanglement.

It is possible to obtain two EPR pairs by applying some two-qubit
unitary operation to an inseparable quantum channel, which enable a
series teleportation of entanglement with rather simple Bell-state
measurement \cite{Note2} and local unitary operation. Unless a quantum
channel is likely to suffer from a reservoir, it may be the simplest
protocol that employs EPR pairs as a quantum channel.  However, when a
reservoir is present, it is important to study inseparable quantum
channels because some inseparable channel can be robuster against the
decoherence than EPR pairs. It is known that some particular state is
robust against the decoherence once the interaction with a reservoir is
known. For example, the decoherence-free state, an eigenstate with zero
eigenvalue of the interaction Hamiltonian, is never decohered in the
given reservoir. We will not further discuss the effects of the
decoherence, which is beyond the scope of this paper.

\section{\bf many-qubit entanglement of inseparable quantum channel}

Any quantum channel in a maximally entangled state of two parts $A$ and
$B$ can be employed for a perfect teleportation of entanglement.
Entanglement of four qubits may be classified into two-qubit
entanglement, tree-qubit entanglement, and four-qubit entanglement.  A
state of four qubits is said to have two-qubit entanglement when some
two qubits among the four qubits are in an entangled state, three-qubit
entanglement when some three qubits are in a three-qubit GHZ state, and
four-qubit entanglement when the four qubits are in a four-qubit GHZ
state.  Note that W-class states and biseparable states \cite{Dur00}
belong to two-qubit entanglement by our definition. As shown in
Sec.~\ref{sec:mqt} a four-qubit GHZ state is not a good candidate for a
perfect teleportation of entanglement. 

The entanglement structure of a possible quantum channel state
(\ref{eq:meqc}) depends on two-qubit unitary operator
$\hat{V}\hat{U}^T$. A quantum channel of two EPR pairs is in the state
$|\bar{\phi}_c\rangle_{AB}$, which is separable in $(A_1,B_1)$ and
$(A_2,B_2)$, and it has only two-qubit entanglement.  We shall present
an example of inseparable quantum channel which has many-qubit
entanglement; the channel state is written as
\begin{eqnarray}
  \label{eq:nbcs}
  |\phi_c\rangle_{AB} &=& \frac{1}{2\sqrt{2}} \bigg( |0000\rangle -
  |0011\rangle + |0101\rangle - |0110\rangle \nonumber \\
  && + |1001\rangle + |1010\rangle + |1100\rangle + |1111\rangle
  \bigg)_{A_1A_2B_1B_2}.
\end{eqnarray}
This state is obtained from Eq.~(\ref{eq:mqc2}) with the two-qubit
unitary operator,
\begin{eqnarray}
  \label{eq:nobe}
  \hat{V}\hat{U}^T = \frac{1}{\sqrt{2}}
  \left(
    \begin{matrix}
      1&0&0&-1\cr 0&1&-1&0\cr 0&1&-1&0\cr 1&0&0&1\cr
    \end{matrix}
  \right),
\end{eqnarray}
which is represented in a product basis set of $\{|00\rangle,
|01\rangle, |10\rangle, |11\rangle\}$. The operator $\hat{V}\hat{U}^T$
transforms the product bases to Bell bases. 

The reduced density operator of each qubit $i$ is proportional to an
identity operator $\rho_i= \openone_i/2$. Noting $|\phi_c\rangle$ is
pure, this implies that $|\phi_c\rangle$ has no individual information
but it contains entanglement of a given qubit and the rest.

To investigate two-qubit entanglement, we employ the Peres-Horodecki
criterion \cite{Peres96,Lee00} for two qubits that their density
operator $\rho$ is entangled if and only if its partial transposition
has any negative eigenvalue. The partial transposition of $\rho$ is
defined as
\begin{eqnarray}
  \rho^{T_{1}}=\sum_{ijkl}\rho_{jikl}|i \rangle\langle j| \otimes |k
  \rangle\langle l|
\end{eqnarray}
when $\rho=\sum_{ijkl}\rho_{ijkl} |i \rangle\langle j| \otimes |k
\rangle\langle l|$. As an example, consider a reduced density operator
of a pair among four qubits which is in a symmetric W-state,
\begin{eqnarray}
 |W\rangle=\frac{1}{2}
(|0001\rangle+|0010\rangle+|0100\rangle+|1000\rangle). 
\end{eqnarray}
The partial transposition of the reduced density operator has a negative
eigenvalue $(1-\sqrt{2})/4$ for all pairs.  

We shall show below that all pairs, which can be selected out of four
qubits in the state $|\phi_c\rangle$, are in separable states. Every
pair $(i,j)$ except $(A_1,B_1)$ and $(A_2,B_2)$ has the reduced density
operator $\rho_{ij}=\frac{1}{4}\openone_{ij}$ and it is disentangled. In
addition, the reduced density operator of $(A_1,B_1)$ or $(A_2, B_2)$ is
given as
\begin{eqnarray}
  \label{eq:scdm}
  \rho_{A_1B_1(A_2B_2)} = \frac{1}{4}
  \left(
    \begin{matrix}
      1&0&0&1\cr 0&1&1&0\cr 0&1&1&0\cr 1&0&0&1\cr
    \end{matrix}
  \right).
\end{eqnarray}
The partial transposition of $\rho_{A_1B_1(A_2B_2)}$ has only positive
eigenvalues of $(0,0,1/2,1/2)$. These results imply that the state
$|\phi_c\rangle$ in Eq.~(\ref{eq:nbcs}) has no two-qubit entanglement.
However, the state $|\phi_c\rangle$ is entangled as shown in the
consideration of the reduced density operators for single qubits and it
has three-qubit entanglement.

\begin{table}[thbp]
  \caption{\label{tab:tghzs}Amplitudes $\lambda_i$ of two orthogonal GHZ states
    $|\phi_0\rangle$ and $|\phi_1\rangle$ in Eq.~(\ref{eq:ets}).}  
\begin{ruledtabular}
\begin{tabular}{crrrr}
    triad & $\lambda_1$ & $\lambda_2$ & $\lambda_3$ & $\lambda_4$ \\
\hline
    $(A_1,A_2,B_1)$ & $-1$ & $1$  & $-1$ & $1$ \\
    $(A_1,A_2,B_2)$ & $1$  & $1$  & $-1$ & $-1$ \\
    $(A_1,B_1,B_2)$ & $-1$ & $1$  & $1$  & $-1$ \\
    $(A_2,B_1,B_2)$ & $-1$ & $-1$ & $1$  & $1$  \\
\end{tabular}
\end{ruledtabular}
\end{table}

A reduced density operator $\rho$ of each triad is obtained by tracing
over the other qubit and it is in the form of
\begin{eqnarray}
  \label{eq:ets}
  \rho = \frac{1}{2} |\phi_0\rangle \langle \phi_0| +
  \frac{1}{2} |\phi_1\rangle \langle \phi_1|
\end{eqnarray}
where $|\phi_0\rangle = |000\rangle + \lambda_1 |011\rangle +
|101\rangle + \lambda_2 |110\rangle$ and $|\phi_1\rangle = \lambda_3
|001\rangle + \lambda_4 |010\rangle + |100\rangle + |111\rangle$ with
$\lambda_i$ given in Tab.~\ref{tab:tghzs}. By generalized Schmidt
decomposition \cite{Thapliyal99} it is found that both $|\phi_0\rangle$
and $|\phi_1\rangle$ are maximal three-qubit GHZ states.

To investigate three-qubit entanglement explicitly, one may employ an
entanglement witness scheme that a density operator of three qubits
$\rho$ has three-qubit entanglement if $\mbox{Tr}({\cal W} \rho) < 0$
for some three-qubit GHZ entanglement witness ${\cal W}$ \cite{Acin01}.
However, it is nontrivial to find such an entanglement witness for a
given density operator while a typical entanglement witness is known as
\cite{Acin01}
\begin{eqnarray}
  \label{eq:ghzw}
  {\cal W} = \frac{3}{4} \openone - |\phi\rangle\langle\phi| 
\end{eqnarray}
where $|\phi\rangle=\frac{1}{\sqrt{2}}\sum_{i=0}^1 |\alpha_i, \beta_i,
\gamma_i\rangle$ is a maximal three-qubit GHZ state. We perform
numerical calculations with steepest decent method to search some local
trilateral rotation for the typical witness (\ref{eq:ghzw}) to minimize
$\mbox{Tr}({\cal W} \rho)$ and we find $\mbox{Tr}({\cal W} \rho) \ge
1/4$. It implies that the typical entanglement witness can not detect
three-qubit entanglement of triads.

\section{\bf Remarks}

We proposed the scheme for entanglement teleportation of a completely
unknown state so as to incorporate a multipartite entangled state as the
quantum channel. Deriving the relations of separabilities for joint
measurements and for corresponding unitary operations, it was found that
an inseparable quantum channel always leads to a total teleportation of
entanglement.  We gave the example of inseparable quantum channel with
each triad in three-qubit entanglement.

\acknowledgments

We would like to thank M.S. Kim, D. Ahn, H.W. Lee, and W. Son for
stimulating discussions.  This work was supported in part by the UK
Engineering and Physical Science Research Council through the Grant
No.~GR/R33304, the Korean Ministry of Science and Technology through the
Creative Research Initiatives Program under Contract
No.~00-C-CT-01-C-35, and supported from the Korean Ministry of Science
and Technology through the National R\&D Program
M10022010001-01G050600110.

\end{document}